# Rules for biological regulation based on error minimization

Guy Shinar*†, Erez Dekel*†, Tsvi Tlusty†, and Uri Alon*†‡

Departments of *Molecular Cell Biology and †Physics of Complex Systems, Weizmann Institute of Science, Rehovot 76100, Israel

The control of gene expression involves complex mechanisms that show large variation in design. For example, genes can be turned on either by the binding of an activator (positive control) or the unbinding of a repressor (negative control). What determines the choice of mode of control for each gene? This study proposes rules for gene regulation based on the assumption that free regulatory sites are exposed to nonspecific binding errors, whereas sites bound to their cognate regulators are protected from errors. Hence, the selected mechanisms keep the sites bound to their designated regulators for most of the time, thus minimizing fitness-reducing errors. This offers an explanation of the empirically demonstrated Savageau demand rule: Genes that are needed often in the natural environment tend to be regulated by activators, and rarely needed genes tend to be regulated by repressors; in both cases, sites are bound for most of the time, and errors are minimized. The fitness advantage of error minimization appears to be readily selectable. The present approach can also generate rules for multi-regulator systems. The error-minimization framework raises several experimentally testable hypotheses. It may also apply to other biological regulation systems, such as those involving protein–protein interactions.

biological physics | complex networks | systems biology | transcriptional regulation

**B**iological regulation systems convert input signals into specified outputs. The same input–output relationship can generally be carried out by several different mechanisms. For example, transcription regulation is carried out by regulatory proteins that bind specific sites in the promoter region of the regulated genes. A gene that is fully expressed only in the presence of a signal (Fig. 1) can be regulated by two different mechanisms (1). In the first mechanism, called positive control, an activator binds the promoter to turn on expression. In the second mechanism, called negative control, a repressor binds the promoter to turn expression off. These two mechanisms realize the same input–output relationship: Expression is turned on by the binding of an activator in the positive mode of control and by the unbinding of a repressor in the negative mode of control. More generally, a gene controlled by $N$ regulators, each of which can be either an activator or a repressor, has $2^N$ possible mechanisms for a given input–output mapping.

Among these equivalent mechanisms, evolutionary selection chooses one for each system. Are there rules that govern the selection of mechanisms in biological systems? One possibility is that evolution chooses randomly between equivalent designs. Hence, the selected mechanism is determined by historical precedent. Another possibility is that general principles exist that govern the choice of mechanism in each system.

The question of rules for gene regulation was raised by M. A. Savageau (2–6) in his pioneering study of transcriptional control. Savageau found that the mode of control is correlated with the *demand*, defined as the fraction of time in the natural environment that the gene product is needed near the high end of its regulatory range. High-demand genes, in which the gene product is required most of the time, tend to have positive (activator) control. Low-demand genes, in which the gene product is not required most of the time, tend to have negative (repressor) control. This demand rule appears to be in agreement with >100 gene systems (2–6) from *Escherichia coli* and other organisms, where the mode of control is known and the demand can be evaluated. For example, the demand rule is in agreement with several dozen well characterized catabolic systems that degrade an externally supplied nutrient. These systems have positive control if the nutrient that they use is found commonly in the natural environment, placing the system in high demand (3, 5). They tend to have negative control if the nutrient is rarely found in the environment, corresponding to low demand for the system. The rule also successfully predicts that systems with antagonistic functions, such as biosynthesis and degradation of a compound, tend to have opposite modes of regulation, whereas systems with aligned functions, such as transport and utilization of a compound, have the same regulation mode (3, 5). This correlation holds regardless of the demand for the system.§

Savageau (3, 5) deduced the demand rule by assuming that repressors and activators are functionally equivalent in all respects except with regard to the effect of mutations. Briefly, high-demand genes have to be expressed most of the time. For such genes, a repressor would accumulate mutations because most mutations tend to eliminate the repressor and thus to keep the gene expressed all of the time. Therefore, there would be little selection pressure against the mutants. Positive regulation, on the other hand, has a strong selection pressure against mutants with eliminated activator because these mutants have deleterious low expression. Hence, a positive mode is more stable against mutations than a negative mode in the case of high-demand genes. A similar argument suggests that in the case of low-demand genes, negative regulation is more stable against mutations than positive regulation.

Such mutant-selection arguments are generally valid only if there is no inherent fitness advantage to one of the two modes of control. If such an inherent fitness difference exists, it would dominate over effects that operate only on mutants.

Here we propose a framework for deducing demand rules for gene regulation based on inherent fitness differences between the two modes of control. The main assumption is that in many regulatory systems, free sites are more error-prone than sites bound to their cognate partner, in the sense that free sites are exposed to binding by nonspecific factors. These errors lead to gene-expression changes, which reduce the organism's fitness. We propose that to minimize fitness-reducing errors, such systems tend to evolve positive control in high-demand environments and negative control in low-demand environments, because in both of these cases the regulatory sites are bound and protected from errors most of the time. We extend our analysis to propose new rules for multiregulator systems, using the *lac* system of *E. coli* as an example, and explain several of its structural features. We also discuss the criteria in which selection according to these rules dominates over historical precedent.

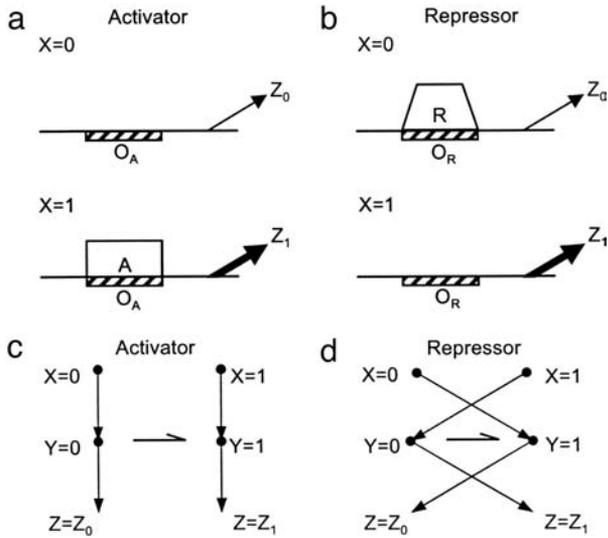

**Fig. 1.** Transcription regulation mechanisms of a gene Z. (*a*) Positive control (activator). In the absence of an inducing signal (input state $X = 0$), the binding site $O_A$ of activator A is free. This causes the gene to be expressed at a low level $Z_0$. When the signal is present ($X = 1$), $O_A$ is bound by the activator A, which causes the gene to be fully expressed ($Z = Z_1$). (*b*) Negative control (repressor). When the inducer signal is absent ($X = 0$), repressor binding site $O_R$ is bound by repressor R. This causes the gene to be expressed at a low level $Z_0$. When the inducer is present ($X = 1$), $O_R$ is free, which causes the gene to be fully expressed ($Z = Z_1$). (*c*) Mapping between input states X, binding states Y, and outputs states Z in the case of positive control. $Y = 0/1$ denotes whether the binding site is free/bound, respectively. The arrow with asymmetric head denotes errors, which affect the free ($Y = 0$) binding state. (*d*) Mapping between input states, binding states, and outputs in the case of negative control.

**Results**

We choose transcription control for the present study, because it is perhaps the best-characterized biological regulation mechanism. Consider first a gene regulated by a transcription factor, which can be an activator or a repressor (Fig. 1). In either case, there is one state in which the regulator binds its site tightly, and another state in which the site is free. This is the idealized model. In reality, the system is embedded in the cell, where many additional regulators and other factors are present. When the site is tightly bound by its designated regulator, the site is protected from these factors. In contrast, when the site is free, it is exposed to nonspecific binding. This nonspecific binding can lead to errors in the expression level of the gene and thus to a reduction in the fitness of the organism.¶ The relative reduction in fitness is called the *error-load*.

There are at least three sources of errors connected with nonspecific binding to the free site. The first source of errors is cross-talk with the other transcription regulators in the cell (10–13). This cross-talk is difficult to prevent, because the concentration and activity of the regulators in the cell changes in response to varying conditions, leading to an ever-changing set of cross-reacting affinities to the site. This cross-talk can act to reduce or increase expression, leading to errors. Because of the dynamic changes in the cross-reactive affinities, it is not possible to cancel the errors out by a basal level of expression. Secondly, lateral gene transfer (14, 15) can introduce exogenous regulatory proteins that can interact with the site, especially when the site is free from its cognate interaction partner. A final source of error arises from residual binding of the designated regulator in its inactive form to its own site: in many cases, the affinity of the inactive regulator is only about one to two orders of magnitude lower than its affinity in the active state (1). Regulator levels are known to fluctuate from cell to cell (16–23), leading to a varying degree of residual binding to the free site, causing cell–cell fluctuations in expression. These errors in expression deviate from the optimal level (7), leading to a reduction in fitness.

We now compare the error-loads of the positive and negative modes of regulation.‖ Consider a gene regulated by an activator, and the same gene regulated by a repressor, such that the two regulatory mechanisms lead to the same input–output relationship. The regulated gene has a demand $p$, defined as the fraction of time that full expression of the gene product is needed in the environment. For both modes of control, errors occur when the DNA site is free and exposed. The two modes differ in the expression state associated with a free site. In the case of a repressor, a free site corresponds to full expression, which occurs a fraction $p$ of the time. Errors in expression lead to a relative fitness reduction of $\Delta f_1$, where the subscript 1 denotes the high expression state. The average reduction in fitness for a repressor, taking into account only errors from the free site, is therefore

$$E_R = p\Delta f_1. \quad [1]$$

For an activator, the free site corresponds to low expression, which occurs a fraction $1 - p$ of the time (the fraction of time that the gene is not in demand). Errors in the expression level lead to a relative fitness reduction of $\Delta f_0$, where the subscript 0 denotes the low expression state. The average reduction in fitness for an activator, taking into account only errors from the free site, is therefore

$$E_A = (1 - p)\Delta f_0. \quad [2]$$

In this simplest case, a repressor will have a fitness advantage over an otherwise equivalent activator when it has a lower error-load.

$$E_R < E_A \quad [3]$$

Using Eqs. **1–3**, we see that repressors are advantageous when the demand is lower than a threshold determined by the ratio of the relative reductions in fitness:

$$p < 1/(1 + \Delta f_1/\Delta f_0). \quad [4]$$

Thus, repressors are advantageous for low-demand genes, and activators are advantageous for high-demand genes (Fig. 2). The reason for this is that repressors in low-demand genes and activators in high-demand genes ensure that the site is bound to its designated regulator most of the time, reducing the fraction of time that the site is exposed to errors.

---

¶Fitness in experiments on growing microorganisms is generally identified with the growth rate (7–9); in other cases, fitness can be identified with the selection coefficient, which is defined as the mean number of offspring per generation.

‖The regulators are assumed to be equivalent in terms of design criteria such as the sharpness of the response function, its dynamic range, etc., as well as the production cost of the regulatory proteins themselves (24). Indeed, sharp and high gain switches are known with both activators and repressors.

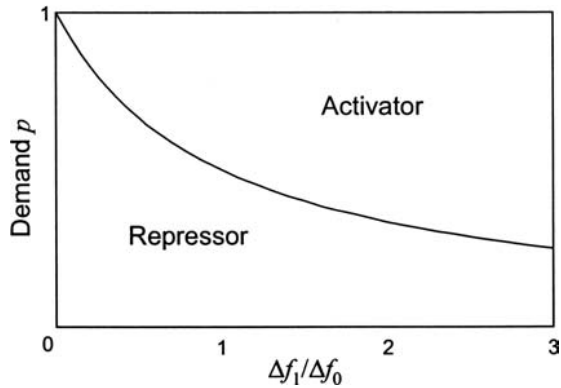

**Fig. 2.** Selection diagram for error-load minimization. Each region corresponds to the mode of control with the smaller error-load. The vertical axis is the demand $p$, defined as the fraction of the time that the gene product is needed at full expression. The horizontal axis is the ratio $\Delta f_1/\Delta f_0$ of the fitness reductions arising from errors in the free sites of the negative and positive control mechanisms.

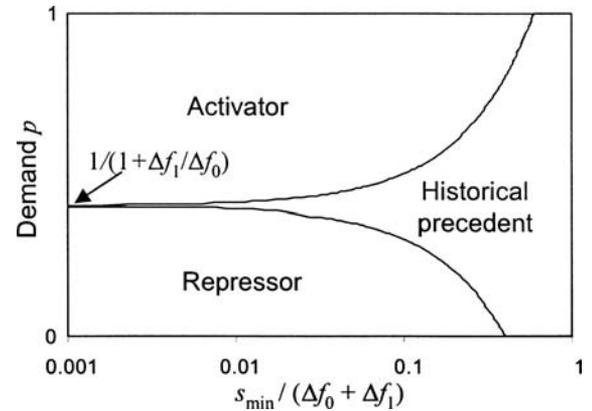

**Fig. 3.** Selectability of error-minimizing mode of control in the face of an existing mechanism with the opposite mode. The region marked "Activator" ("Repressor") is a region in which a mutant bearing an activator (repressor) regulatory system can become fixed in a wild-type population carrying the opposite mode of control. In the region marked "Historical precedent," mutants with optimal regulatory mechanisms do not have sufficient fitness advantage to take over a population with an existing, suboptimal mode of regulation. The $x$ axis is $s_{min}/(\Delta f_0 + \Delta f_1)$, where $s_{min}$ is the minimal selection advantage needed for fixation, $\Delta f_0$ is the reduction in fitness due to errors in the free state of an activator site, and $\Delta f_1$ is the reduction in fitness due to errors in the free state of a repressor site. In this plot, the ratio $\Delta f_1/\Delta f_0$ is constant, equal to 1.5.

Can error-load create a selection pressure sufficient to cause a regulatory system to be replaced by a system with the opposite mode of control? Consider a wild-type population with a regulatory system in place. Suppose that conditions vary, leading to a permanent change in the demand for the gene, so that the opposite mode of control becomes optimal. Mutants with the opposite mode of control arise in the population, by genomic mutation or lateral gene transfer.** These mutants have a lower error-load, and hence a relative fitness advantage, which is equal to $E_A - E_R$ in the case of a repressor and $E_R - E_A$ in the case of an activator. The mutants can become fixed if their relative fitness advantage exceeds a minimal selection threshold, $s_{min}$. The selection threshold $s_{min}$ has been estimated in bacteria and yeast to be in the range of $10^{-8}$ to $10^{-7}$ (9, 26).

The condition for fixation of a repressor mutant is thus $E_A - E_R > s_{min}$, whereas the condition for fixation of an activator mutant is $E_R - E_A > s_{min}$. These inequalities lead to a selection diagram (Fig. 3), in which the error-minimizing regulatory mechanism becomes fixed at a given demand $p$ only if the ratio $s_{min}/(\Delta f_0 + \Delta f_1)$ is sufficiently small. If this ratio is large, there exists a region in parameter space where historical precedent determines the mode of control.

One can estimate whether expression errors lead to selectable error-load differences. The fitness as a function of protein expression was experimentally determined in the *lac* system of *E. coli* (7). The resulting fitness function indicates that a 1% error in expression leads to $\Delta f_0$ and $\Delta f_1$ on the order of $10^{-3}$, which is four orders of magnitude higher than the selection threshold $s_{min}$. Similarly, Wagner (9) estimated that in yeast, a 2% change in the expression level of any protein is sufficient to cause fitness differences that exceed the selection threshold. These considerations suggest that even minute expression errors lead to error-load effects that can dominate over historical precedent in determining the choice of regulatory system.

So far, we have analyzed the demand rules for a gene with a single regulator. We now turn to systems with multiple regulators. For clarity, we will consider in detail the *lac* system of *E. coli*, though the present considerations can be generally applied to other systems.

---

**There are well characterized examples where the same regulatory protein can act either as a repressor or as an activator depending on the position and strength of its regulatory site (1, 25); other cases are known where mutations in the regulator coding region can cause a repressor to become an activator and vice versa (1).

The Lac proteins transport the sugar lactose into the cell and participate in its degradation for use as a carbon and energy source (1, 27). The system has two input stimuli, lactose and glucose. Expression of the Lac proteins is induced in the presence of lactose to allow utilization of the sugar. Expression is inhibited in the presence of glucose, which is a better energy source than lactose. The input–output relationship, which maps the levels of the two input sugars onto the expression levels of the Lac proteins, is shown in Fig. 4a.

This input–output relationship is implemented by two regulators, the repressor LacI and the activator CRP. When lactose enters the cell, the repressor LacI does not bind its DNA sites, causing increased expression of the Lac proteins. When glucose enters the cells, the activator CRP does not bind its DNA site, leading to a reduction in expression. The *lac* system has an additional mechanism that inhibits expression in the presence of glucose, which is called *inducer-exclusion*: when glucose is pumped into the cell, lactose entry is blocked, preventing the induction of the *lac* system (28, 29).

The relation between the input states, the DNA binding states, and the output levels of the *lac* system is shown in Fig. 4b. There are four possible binding states, depending on whether the CRP and LacI sites are bound or free. These binding states are denoted [CRP,LacI] = [0,0], [0,1], [1,0], and [1,1], where 1/0 correspond to bound/free. One of the four binding states, [CRP, LacI] = [0,0], is not reached by any input state, because inducer exclusion prevents lactose and glucose from being present in the cell at the same time. Thus, in the presence of glucose, the LacI site is bound even if lactose is present in the environment. As a result, the binding-state [0,0] does not correspond to any input state and may therefore be called an *excluded* state. Its expression level was experimentally determined by using artificial inducers such as IPTG, which are not subject to inducer exclusion (27, 30).

The naturally occurring mechanism, with a glucose-responsive activator and a lactose-responsive repressor, is only one of the four possible mechanisms in which the two regulators can have either mode of control (Fig. 5). The four mechanisms can be denoted RR, RA, AR, and AA where the first letter denotes the mode of the glucose regulator, the second letter denotes the mode of the lactose regulator, and A/R corresponds to activator/repressor. The wild-type *lac* system has the AR mechanism, with activator CRP and repressor LacI (Fig. 5*a*).

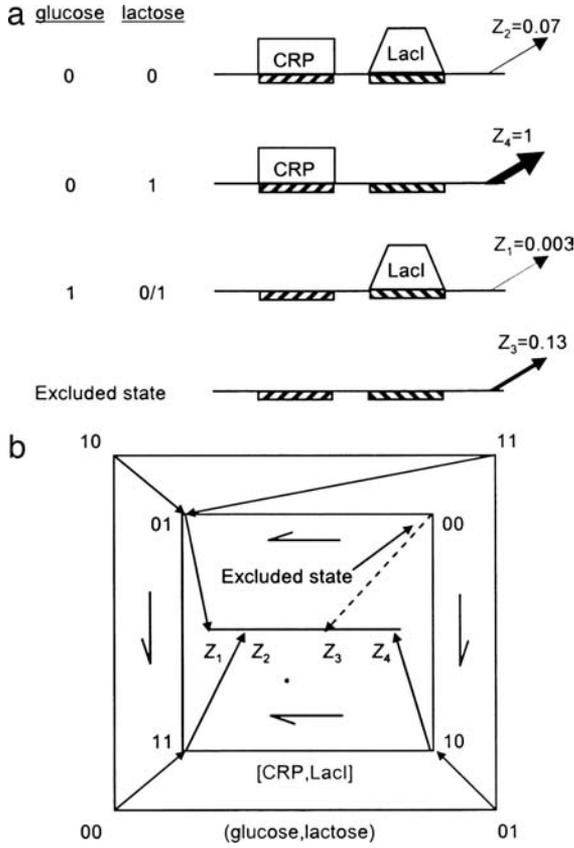

**Fig. 4.** The *lac* system of *E. coli* and its regulatory mechanism. (*a*) Input–output relationship of the *lac* system and its molecular implementation. Glucose and lactose levels are 1/0 corresponding to saturating/no sugar in the environment. $Z_1$, $Z_2$, $Z_3$, and $Z_4$ are the relative protein expression levels from the *lac* promoter (30). The binding states of CRP and LacI are shown. (*b*) Diagram mapping the input states (glucose,lactose) onto the binding states [CRP,LacI] and finally onto the output states $Z_1$, $Z_2$, $Z_3$, and $Z_4$. The dashed arrow indicates the excluded state, which is not reached by any of the input states because of inducer exclusion. The arrows with asymmetric heads indicate errors by pointing away from free binding states.

In all of these mechanisms, the input states map onto the expression levels in the same way. The mechanisms differ in the promoter binding states that correspond to each input and output state. All four mechanisms have inducer exclusion and thus have an excluded state, although the identity of the excluded state differs between the mechanisms (Fig. 5): The excluded state is [CRP,LacI] = [0,0], [0,1], [1,0], and [1,1] in the AR, AA, RR, and RA mechanisms, respectively.

Table 1 lists the fitness reductions resulting from errors that occur when one or both of the regulator sites are free. The rows correspond to the four mechanisms, and the columns correspond to the input states. The reductions in fitness due to errors from the free binding sites of the glucose- and lactose-responsive regulators are $\Delta f_i$ and $\Delta f_i'$, respectively, where the index $i = 1\ldots 4$ denotes the output level that corresponds to each input state. For example, consider the AR mechanism with the input state (glucose,lactose) = (0,1). This input state, which corresponds to the highest expression level $Z_4$, is mapped onto binding state [CRP,LacI] = [1,0], where the glucose-responsive regulator site is bound and the lactose-responsive regulator site is free. Thus,

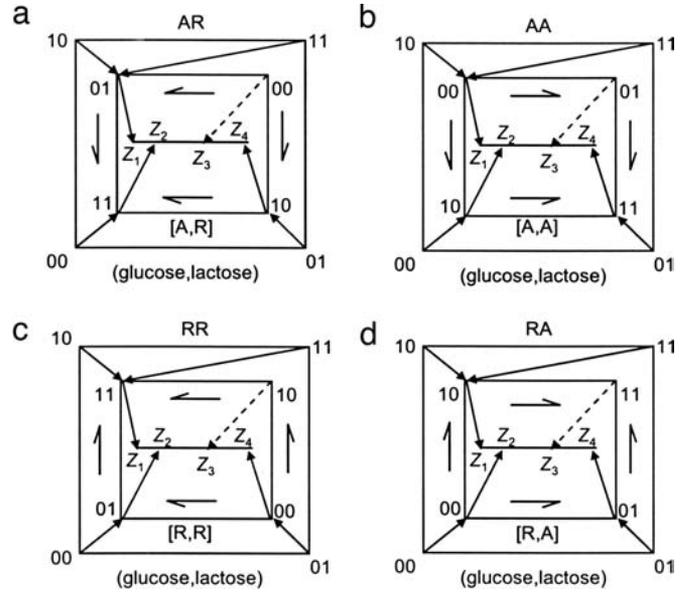

**Fig. 5.** The four possible regulatory mechanisms of the *lac* system. The mechanisms are labeled by the mode of regulation of the glucose- and lactose-responsive regulators, where A/R means activator/repressor.

only the latter site is exposed to errors, contributing a reduction $\Delta f_4'$ to fitness.

The error-load is calculated by multiplying the probability of each input state by the appropriate fitness reductions and summing over all input states (Table 1). For example, the error-load of the AR mechanism is

$$E_{AR} = p_{01}\Delta f_4' + (1 - p_{00} - p_{01})\Delta f_1, \quad [5]$$

where $p_{00}$ is the probability that neither glucose nor lactose is present in the environment and $p_{01}$ is the probability that glucose is absent but lactose is present.

We can now compare the different mechanisms and determine which has the lowest error-load in a given environment ($p_{00}$, $p_{01}$). This comparison results in the selection diagram that appears in Fig. 6. The diagram is triangular because $p_{00} + p_{01} \leq 1$. We find that three of the four mechanisms minimize the error-load, each in a different region of the diagram. One of the four possible mechanisms (RA) never minimizes the error-load. The wild-type mechanism of the *lac* system, AR, minimizes the error-load in a region of the diagram that includes environments where both lactose and glucose are present with low probability, namely $p_{01} \ll 1$ and $p_{00} \approx 1$. This finding is consistent with the empirical observation that in the natural environment of *E. coli*, both glucose and lactose are rare (2, 5, 31).

**Table 1. Fitness reduction in the *lac* system as a function of regulatory mechanism and input state**

| Mechanism/input state | (0,0) | (0,1) | (1,0)/(1,1) |
|---|---|---|---|
| AA | $\Delta f_2$ | 0 | $\Delta f_1 + \Delta f_1'$ |
| AR | 0 | $\Delta f_4'$ | $\Delta f_1$ |
| RA | $\Delta f_2 + \Delta f_2'$ | $\Delta f_4$ | $\Delta f_1'$ |
| RR | $\Delta f_2$ | $\Delta f_4 + \Delta f_4'$ | 0 |

Columns correspond to input states (glucose,lactose), where 1/0 means saturating/no input. Rows correspond to the possible regulatory mechanisms. The reductions in fitness due to errors from the glucose-responsive (lactose-responsive) regulator binding site are $\Delta f_i$ ($\Delta f_i'$), where $i = 1\ldots 4$ corresponds to the output level. The input states (glucose,lactose) = (1,0) or (1,1) are both mapped onto the same binding state because of inducer exclusion.

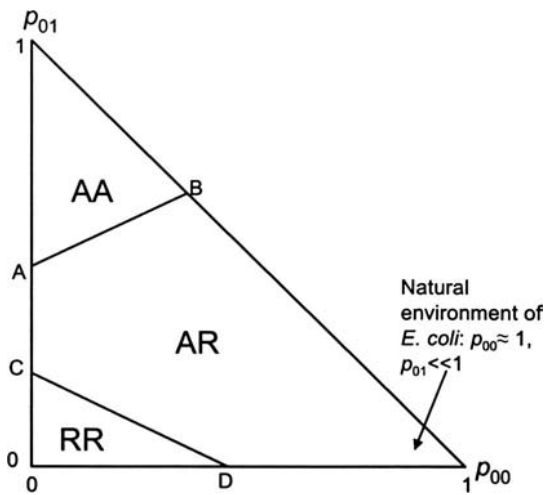

**Fig. 6.** Selection diagram for the *lac* system, indicating the regulatory mechanism that minimizes the error-load in each environment. The axes are the probability for neither glucose nor lactose in the environment ($p_{00}$), and for lactose in the absence of glucose ($p_{01}$). The wild-type mechanism, AR, minimizes the error-load in a region that includes environments where both sugars are rare. Line AB has slope $(\Delta f_2' - \Delta f_1')/(\Delta f_1' + \Delta f_4')$ and intercepts the vertical axis at $\Delta f_1'/(\Delta f_1' + \Delta f_4')$. Line CD has slope $-(\Delta f_1 + \Delta f_2)/(\Delta f_1 + \Delta f_4)$ and intercepts the vertical axis at $\Delta f_1/(\Delta f_1 + \Delta f_4)$.

It is easy to see why the AR mechanism minimizes the error-load in the natural environment of *E. coli*. The most frequent input state is (glucose,lactose) = (0,0), which maps onto the second lowest output level $Z_2$. In the AR mechanism, this corresponds to the binding state [CRP,LacI] = [1,1], where both regulators bind their DNA sites. Thus, the AR mechanism keeps the DNA sites protected from errors most of the time. In addition, the AR mechanism has another error-minimizing feature: the most error-prone binding state [CRP,LacI] = [0,0], in which both sites are free, is concealed by inducer exclusion. Hence, not only does the AR mechanism map the most frequent input state onto the error-free binding state [1,1], but it also excludes the most error-prone binding state [0,0] and prevents it from being reached by any input state.

**Discussion**

This study suggests that biological regulation systems in which open sites are error-prone will tend to evolve mechanisms that keep the sites bound for most of the time, thus minimizing errors. In the case of transcription regulation, genes whose product is required at full expression for a small fraction of the time (low-demand genes) will tend to have repressor control. Genes needed at full expression a large fraction of the time (high-demand genes) will tend to have activator control. This approach is easily generalized to multi-input systems, as demonstrated for the *E. coli lac* system. The expected selective advantage of error-load minimization in transcription regulation appears to be sufficient to overcome historical precedent in the choice of regulatory mechanism.

The present conclusions for transcription regulation assume that free DNA sites are more error-prone than sites bound to their cognate regulators. In cases where the reverse is true, that is when bound sites are more error-prone, the predictions are opposite, namely that repressors (activators) correspond to high (low) demand genes. One possible scenario in which bound sites might be more error-prone than free sites may occur in eukaryotic genes in which unbound sites are tightly packed in closed chromatin conformations. In this case, closed chromatin may protect the free sites from errors, whereas regulator binding may require opening of chromatin allowing increased exposure to errors (1, 21, 22, 32). Further study is needed to assess the error-loads associated with such chromatin states.

The present study raises several experimentally testable hypotheses. First, it raises the challenge of directly measuring the magnitude of expression errors from a free regulatory site as compared with the same site bound by its cognate regulator. One possibility is to measure expression from a promoter with a given regulatory site in individual cells (16–23), while manipulating the intracellular activity of regulators that recognize similar sites. Second, it is possible to compare the fitness associated with alternative regulatory mechanisms. One way to do this is to construct the desired regulation mechanism out of well characterized components [a synthetic circuit approach (23, 33–41)] and then to compare the fitness of strains bearing different mechanisms by means of direct competition experiments (8). Finally, evolution experiments in defined environments (7, 8, 42) that allow control of the fraction of time where each input is found can test whether the predicted regulation mechanisms actually evolve in each environment, and on what time scale.

Although this study addressed transcription regulation, we note that the present approach can be applied to other biological systems in which regulation involves the binding of biomolecules. One example is protein–protein interactions mediated by specific binding domains (43, 44). In this case, positive and negative modes of regulation correspond to the activation of proteins by either the binding or unbinding of a regulatory domain. Experiments by Lim and colleagues (43) have indicated that cross-talk between different SH3 domains in the same cell sets a selectable constraint on the fitness of the organism. Selection of mode of control (positive or negative) according to demand rules is a possible way for the cells to evolve increased specificity, by minimizing the time that a given site is free and exposed to cross-reactivity. This hypothesis has experimentally testable predictions in the case of protein–protein interactions, in the same spirit as those suggested above for transcription control.

In summary, this study proposed rules for biological regulation based on minimizing error-load. In systems where a free site is more exposed to errors than a site bound to its cognate interaction partner, it is predicted that mechanisms that keep the site bound most of the time will have a lower error-load and hence a selectable advantage. In the context of transcription, this explains the Savageau demand rule and offers rules for multi-regulator systems. Future experiments and theoretical studies can test the generality of these rules, as well as their applicability to other modes of biological regulation.

We thank M. A. Savageau for important discussions; M. B. Elowitz, E. Hornstein, and Z. Kam for comments on the manuscript; the members of our laboratory for helpful input; and the faculty and students of the 60th Cold Spring Harbor Bacterial Genetics course for important discussions. This work was supported by the Israel Science Foundation, the National Institutes of Health, the Human Frontier Science Program, and the Kahn Family Foundation.